\documentclass[preprint,10pt]{elsarticle}

\usepackage{amssymb}
\usepackage{amsfonts}
\usepackage{amsmath}
\usepackage{color}
\usepackage{booktabs}
\usepackage{mathbbol}
\usepackage{mathrsfs}
\usepackage{epsfig,bm,dcolumn}
\usepackage{graphicx}
\usepackage{geometry}
\usepackage{overpic}
\usepackage{slashed}
\usepackage[bookmarks,bookmarksnumbered]{hyperref}
\hypersetup{colorlinks = true,linkcolor = blue,anchorcolor =red,citecolor = blue,filecolor = red,urlcolor = red,
            pdfauthor=author}
\usepackage[capitalise]{cleveref}
\usepackage{txfonts}
\usepackage{floatrow}
\usepackage{subfig}
\journal{XXX}

\begin{document}

\begin{frontmatter}



\title{Measuring the rationality in evacuation behavior with deep learning}

\author[us]{Huaidian Hou}
\address[us]{The Haverford School, 450 Lancaster Avenue, Haverford, PA 19010, United States}

\author[fias]{Lingxiao Wang\corref{cor}}
\address[fias]{Frankfurt Institute for Advanced Studies, Ruth-Moufang-Str. 1, 60438 Frankfurt am Main, Germany}
\cortext[cor]{Corresponding author.}

\begin{abstract}
The bounded rationality is a crucial component in human behaviors. It plays a key role in the typical collective behavior of evacuation, in which the heterogeneous information leads to the deviation of rational choices. In this study, we propose a deep learning framework to extract the quantitative deviation which emerges in a cellular automaton(CA) model describing the evacuation. The well-trained deep convolutional neural networks(CNNs) accurately predict the rational factors from multi-frame images generated by the CA model. In addition, it should be noted that the performance of this machine is robust to the incomplete images corresponding to global information loss. Moreover, this framework provides us with a playground in which the rationality is measured in evacuation and the scheme could also be generalized to other well-designed virtual experiments.
\end{abstract}


\end{frontmatter}

\section{Introduction}
\label{sec: intro}

As one of collective behaviors under extreme condition, crowd congestion in emergencies routinely causes disaster, such as clogging and stampede~\cite{helbing:2000simulating,hughes:2002continuum,helbing:2005selforganized}. It becomes significant to investigate the collective patterns and behaviors in case of emergency. Moreover, the relevant researches could help bridge the cognitive gap between individual decisions and collective behaviors under extreme condition. The collective patterns have been demonstrating in simulations and experiments where the group movement behaves as short intermittent bursts~\cite{pastor:2015experimental,nicolas:2018counterintuitive,bain:2019dynamic}. With the desire of escaping danger exceeding the instinct of avoiding collisions, pedestrian's behavior transforms from order to disorder. On the individual decision side, game theory, decision theory, communication model and queuing model have been comprehensively applied in evacuation process~\cite{vermuyten:2016review}. 

The simulation researches mainly concentrate in macroscopic collective models and microscopic behavior models. Considering individual self-organization emerges from inner interactions, there were works applying hydrodynamic model to study the macroscopic collective behavior of the population~\cite{low:2000statistical,hughes:2002continuum}, and the afterwards researches revealed more non-trivial macroscopic patterns~\cite{bain:2019dynamic,nicolas:2019mechanical,ma:2021spontaneous}. 
With regard to individual strategies, the micro-models represented by the Social force model, the cellular automaton(CA) model, and the magnetic field force model~\cite{helbing:1995social,helbing:2000simulating,burstedde:2001simulation,weng:2006cellular,guo:2012heterogeneous} could give a more completed description of the evacuation behavior. The simulating results are also validated in experiments, such as the "faster-is slower" phenomena were observed in experiments containing different individuals, e.g, people, vehicles, ants, sheep, microbial populations, and so on~\cite{pastor:2015experimental,patterson:2017clogging,aguilar:2018collective,delarue:2016selfdriven,garcimartin:2015flow}. 
The evacuation simulation provides us with a platform in which the human instinct dominates the behaviors under an extreme scene~\cite{nowak:2005emergence,zanlungo:2017intrinsic,corbetta:2017fluctuations,guo:2012heterogeneous,nicolas:2018trap,cavagna:2018physics}.  It brings opportunities to effectively investigate human behavior itself without complex social relations, which will help us to understand the diverse and fascinating collective behaviors occur in both virtual and real space\footnote{Social network, financial network and social norms, these virtual social connections naturally incubate the collective behavior; as for the real space, collective modes are common in urban dynamics, traffic flow, and pedestrian dynamics~\cite{castellano:2009statistical,ball:2012why}.}.

\begin{figure}[htbp!]
\centering\includegraphics[width=0.9\linewidth]{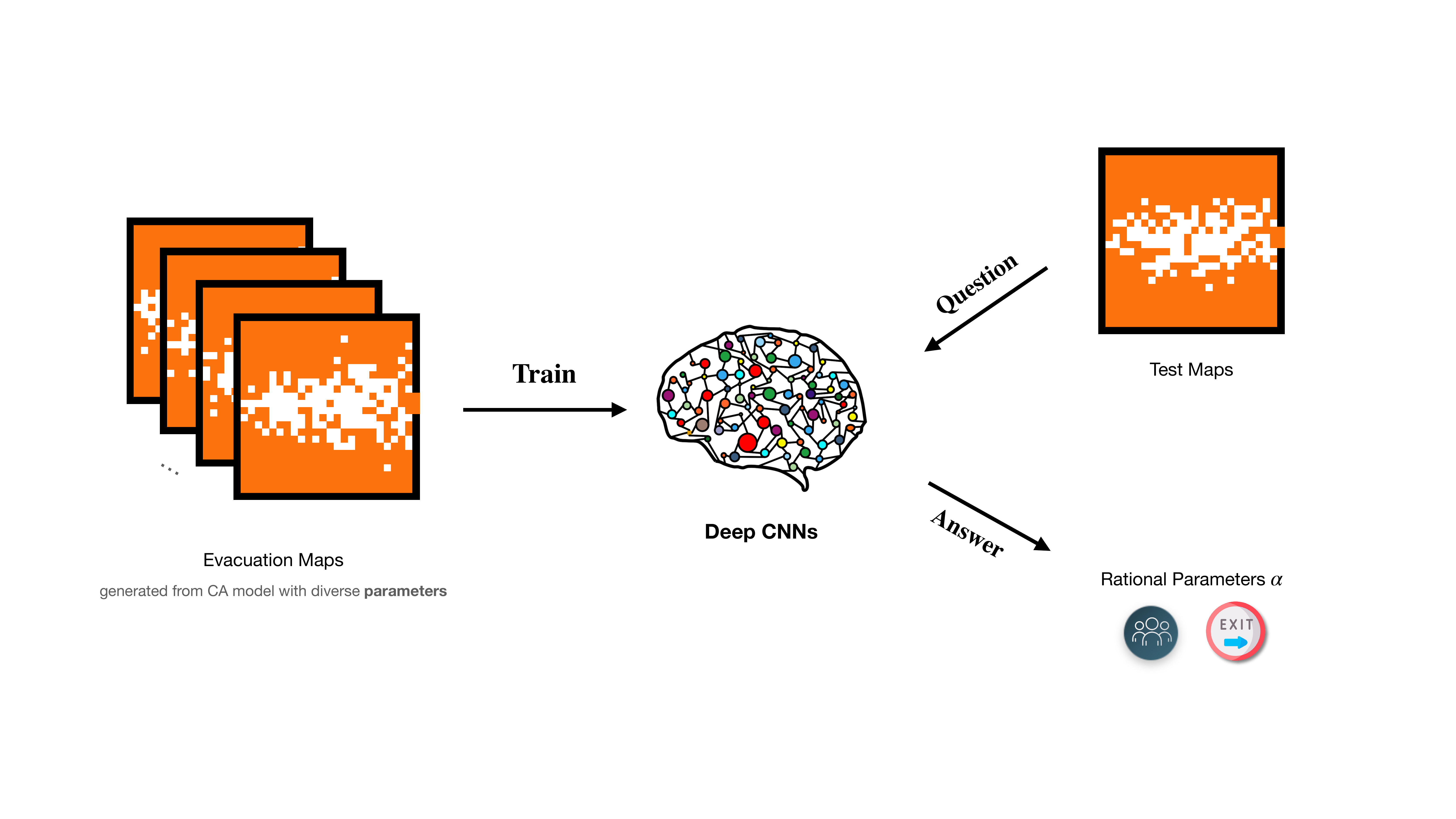}
\caption{Flowchart of learning the dynamical parameters in evacuation models and estimate the values in other cases.}
\label{fig: flow}
\end{figure}

For the past few years, the development of sensor technology and the improvement of microchip computing power have been yielding unusually brilliant results in diverse fields. It makes things feasible that is collecting abundant data and using the state-of-the-art machine learning methods to simulate the escape process ~\cite{moussaid:2011how,helbing:2011recognition,corbetta:2017fluctuations,zanlungo:2017intrinsic,wang:2018study}. Deep learning (DL), a branch of artificial intelligence(AI), efficiently integrates statistical and inference algorithms and thus offers the opportunity to uncover hidden structures of evolution in complex data and to describe it with finite dynamical parameters. Therefore, a combination of DL algorithms and spatio-temporal models for evacuation based on bounded rationality is a promising option. The existing researches mainly focus on applications of DL in designing evacuation strategies based on data~\cite{rahman:2018shortterm,song:2017deepmob,chen:2020application}. The other potential application is to train deep neural networks(DNNs) on simulated data-sets and transfer it into real data-sets to evaluate realistic situations or recognize the hidden signals, which has been verified in both physics and epidemiology~\cite{pang:2018equationofstatemeter,jiang:2021deep,wang:2021machine}. Based on the methodology, as Fig.~\ref{fig: flow} shows, we first introduce the DL into evacuation model to measure the rationality emerges in such extreme scene.

In this paper, we use replicator dynamics to simulate the evacuation, which combines the bounded rational behavior and rational decision-making~\cite{wang:2019escape,pan:2014spatial,heliovaara:2013patient,taylor:1978evolutionary}. Bounded rationality is formalized with such major variables as incomplete information, information processing and the non-traditional policy-maker target function~\cite{simon:1983reason,gigerenzer:2002bounded,yang:2005walrasian,babeanu:2018evidence,pan:2014spatial}. Heterogeneous information could be the reason why people shows irrationality~\cite{lee:2001effects,nowak:2005emergence,wang:2005advanced,moussaid:2011how,wang:2019escape}. Adopting the simulation model proposed in Ref.~\cite{wang:2019escape}, we deploy a DNN model to extract the rational parameters which decide the individual behavior in a Cellular Automaton model describing the evacuation.
To train the deep convolutional neural networks (CNNs), we prepared the data-set with diverse rational factors from multi-frame images generated by the CA model, which specifically means the CNNs are trained on the images cut from whole evolutions. In addition, the framework has been evaluated on 4 different CNN models, and further examined on the data-sets consist of incomplete images which are cropped from original images. This framework provides us with a playground where rationality could be measured in evacuation. It is first validated on the data-set simulated with the $ crowd $ rule~\cite{wang:2019escape} which is a typical behavior with the bounded rationality.

\section{Extracting Rationality in Evacuation Behavior with Deep Learning}
\label{sec: ml}

\subsection{Cellular automaton modeling evacuation with bounded rationality}
A cellular automation model was proposed for simulating the pedestrian flow with bounded rationality in a two-dimensional system~\cite{wang:2019escape}. The underlying structure is a $ L\times L $ cell grid, where $ L $ is the system size. The state of cell can be empty, or occupied by one pedestrian exactly or wall. The Moore neighbor is adopted in the CA model, and pedestrians update their positions by transition matrices $ P(i,t) $, where $ P_{m,n}(i,t) $ means the possibility that the pedestrian $ i $ moves from $ t $ time at position $ (x(i,t),y(i,t) ) $ to next time-step position. The neighbors' directions are labeled by $ (m,n)$, where $ m,n=1,2,3 $.  Each cell could be either empty or occupied by a wall or a pedestrian.  Pedestrians at each time step can choose to move into a new location or stop.  Once we choose the location of the exit, the cellular automata updated synchronously can simulate the escape process. ~\cite{burstedde:2001simulation,kirchner:2003friction}.

The model escape rules gives as follows: Set the position of exit $ (x, y) $ and generate $ N(i,t) $ population distribution at the $ L\times L $ lattice. At the $ t=0 $ time, disaster turns out and individuals begin to move; at the $ t $ time step, the individual $ i $ move to the next position as transition matrices $ T(i,t) $ at $ t+1 $ time step. Update all individuals synchronously, and the conflict will be handled by compared the transition possibility; {\emph{Handle Conflicts}. The conflicts occurs when the two or more persons want to move into the same position, and what we do to handle the conflicts is to compare their transition possibilities $ P_{m,n}(i,t)$ which reflects their willingness to move. For example, the individual $j$ and $k$ both want to move into position $(x,y)$, and the corresponding possibility for the $j$ is  $ P_{m,n}(j,t)$ and the $k$ is $ P_{m',n'}(k,t)$. If the $ P_{m,n}(j,t)>P_{m',n'}(k,t)$, then the individual $j$ move successfully and the $k$ stayed where it was, and vice versa. For equal cases, one is randomly selected. It can be easily extended to the situation of many people. } For the individuals whose destination is exit at the next time step, they escape successfully and are removed from the area to reduce population as $ N(t+1)=N(t)-1 $. If $N(t) = 0$, all individuals exited and stop evolution; Else, update transition matrix according to the above strategies.

The extreme situation of escaping from disasters constrains people's behavior, in which only intuition or social habits remains, no long term trade-off. The replicator dynamics modeling~\cite{taylor:1978evolutionary,heliovaara:2013patient} links the different behaviors, whether practical or spiritual, during the escaping process. It reforms the transition possibility $ P(i,t) $ as,
\begin{equation}
	P_{m,n}(i,t)=\frac{B_{m,n}(i,t)R_{m,n}(i,t)}{\sum B(i,t)R(i,t)}
\end{equation}
where $ R(i,t), B(i,t) $ means the weight from rational and bounded rational part respectively. The definition of the components in matrix $R_{m,n}(i,t)=O_{m,n}(i,t)E_{m,n}(i,t) $,
\begin{equation}
	O_{m,n}(i,t)=\left\{
	\begin{array}{rcl}
		1 &  & {empty} \\ 
		\epsilon &  & {occupied}
	\end{array},  E_{m,n}(i,t)=\Bigg\{\begin{array}{rcl}
		\alpha &  & {exit} \\ 
		\epsilon &  & {nothing}
	\end{array}\right.
\end{equation}
which means if the position $ (m,n) $ around the individual $ i $ at $ t $ time is empty, the $ O_{m,n}(i,t)=1 $, whereas the value is $ \epsilon $. And the $ E_{m,n}(i,t)=\alpha $ only holds when the exit direction is indicated by $ (m,n) $, if not take the value $ \epsilon $. The $ \epsilon $ is a minimum value that the calculation accuracy can reach. The parameter $ \alpha $ represents  the attraction of the exit to persons want to escape, or the importance of the information of the exit position.

The definition of the bounded rational part $ B_{m, n}$ relies on the heterogeneous information from the crowd. As the transport model of statistical physics inspired us, the escape dynamics needs more information on the persons' position and velocity distribution, the basic variables of the transport theory. Considering the full information cannot easily be observed by individuals, the mean-field approximate can provide a global perception for the people on move, which shows $B_{m,n}(i,t)= 1$ as rational choices, $B_{m,n}(i,t) = n_{m,n}(i,t)$ as influencing by the crowds. The $ rational $ indicates the transition possibility only decided by $ R(i,t) $, the neighbor occupied state and the direction of exit, or the objective environment. The $ crowd $ defines $ n_{m,n}(i,t)={\sum_{m,n}N(i, t)}/{\sum_{All} N(i, t)} $, where $ N(i, t) $ is the population distribution at $ t $ time. The definition shows the proportion of individuals in $ (m,n) $ orientation as mean-field approximation, and people will be attracted to the direction with more density. We use it to mimic the "crowd" behavior for individuals, which also means people can potentially get more population density information. The $crowd$ effect induced by population affects the human behavior indirectly, since people can gather and process information from the environment~\cite{moussaid:2011how,lee:2001effects}. In this work, the distribution is discrete and the individual can process them as background, that's what above definition means. People's perception to the distribution is reduced to the average value in a certain direction, a mean background field, as what  statistical physics did in a many-body system.
\begin{figure}[htbp!]
\centering\includegraphics[width=0.8\linewidth]{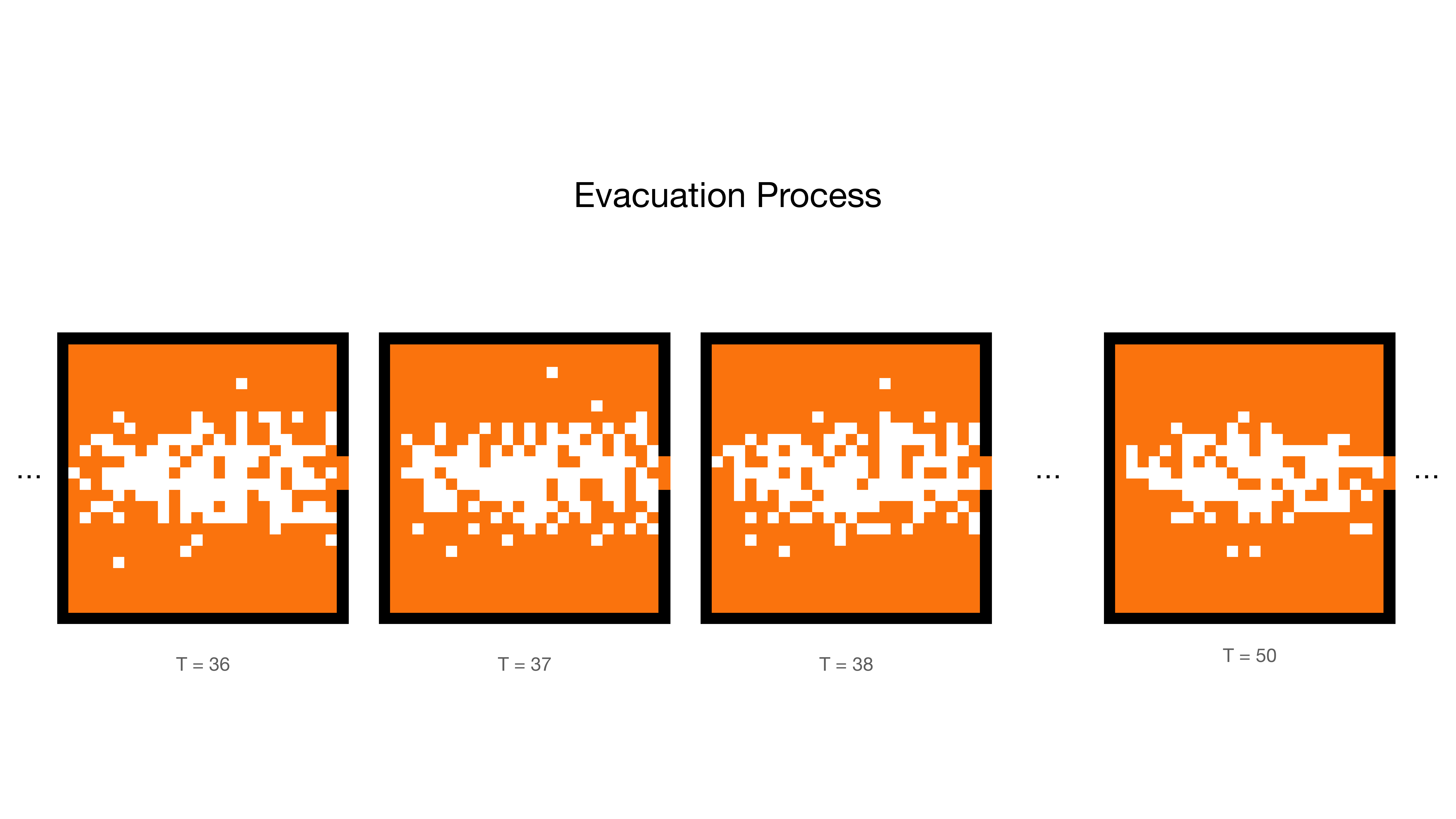}
\caption{The simulation of an evacuation process with parameters $L= 24, \alpha=10, \rho_0=0.37$.}
\label{fig: evacuation}
\end{figure}

\subsection{Data-set generation and network capacity}
The data-sets that we prepared for training the neural networks are from the CA model included a total of 50000 images. Out of the 50000, there are 5 different initial populations $\rho_0$ ranging from 0.1 to 0.5, each with 10000 images generated. Out of the 10000 images with each initial population, there are 100 different values of rational parameters $\alpha \in (0,5)$ and 100 frames of evolution in Time-step $T\in [1,100]$ for each parameter. Each image represents one snapshot of the evacuation process in a square form with a side length of 24, so each image we generated has 576 pixels. Each pixel of an image is either 0 or 1, where 0 represents empty space and 1 represents an individual present at that spot. 

\begin{figure}[h]
\centering\includegraphics[width=1.0\linewidth]{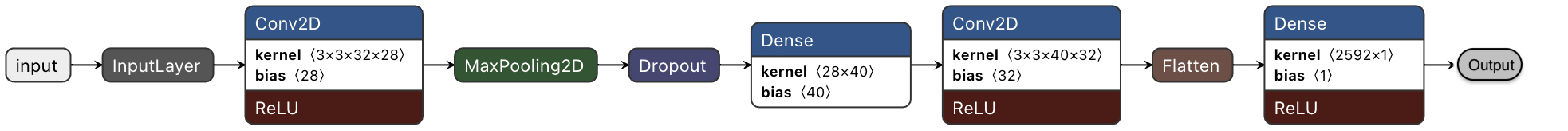}
\caption{The CNN model we used to learn rational parameters from evolution.}
\label{fig: cnn}
\end{figure}
The main architecture of the CNNs we used in this study is shown in Fig.~\ref{fig: cnn}. As the model to process the images generated from the cellular automaton, the Conv2D layer is following after the input layer, and the MaxPooling layer is used to coarsening the features extracted from the CNN. The second Conv2D layer could be expanded to more CNNs whose performance is demonstrated in Sec.~\ref{sec:res}. The fully-connected layers before the output layer is applied to process the signals from the preceding CNNs. The Droupout module and $L_2$ regularization are deployed to alleviate the possible over-fitting. To prepare the inputs for above CNN models, we select 10000 as a standard batch size of samples, in which 2000 samples are from 5 
initial population panels and mixed in one training data-set. Out of the 2000, we labeld all 100 rational parameters $\alpha$ to each frame, and prepared 20 groups from different frame selections. It means we prepare different numbers of consecutive frames as training data-sets, which helps us to evaluate the performance of the CNNs to extract the dynamical information from the collective behavior. Starting with frame No. 36 as Fig.~\ref{fig: evacuation} shown, we cut the following one frame as the first channel of 2000 samples, and then cut different numbers of frames (ranging from 1 to 32) after the first frame to form diverse channels of the image inputs.

\section{Results}
\label{sec:res}
\subsection{Validation of the CNN models}

\begin{figure}[htbp!]
\centering
\sidesubfloat[]{\includegraphics[width=5.0cm]{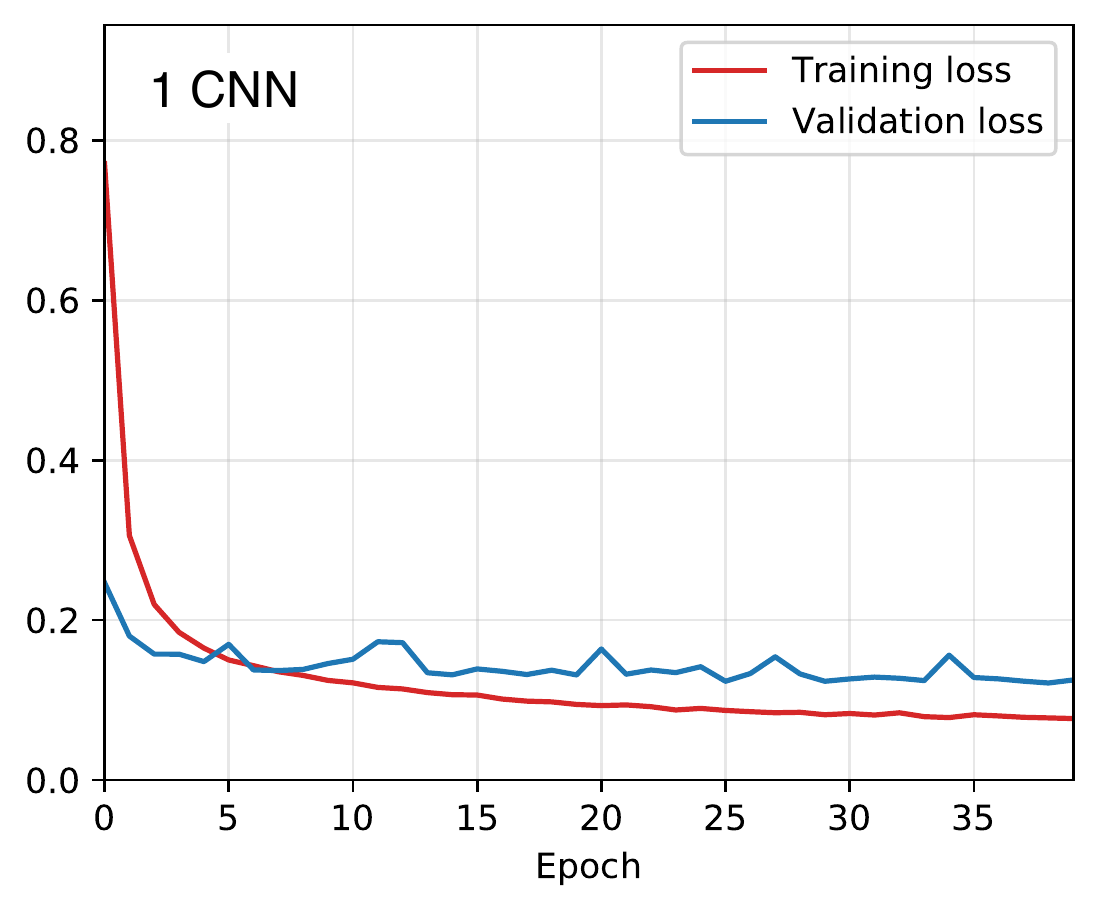}\label{fig:1conv}}
\,
\sidesubfloat[]{\includegraphics[width=5.0cm]{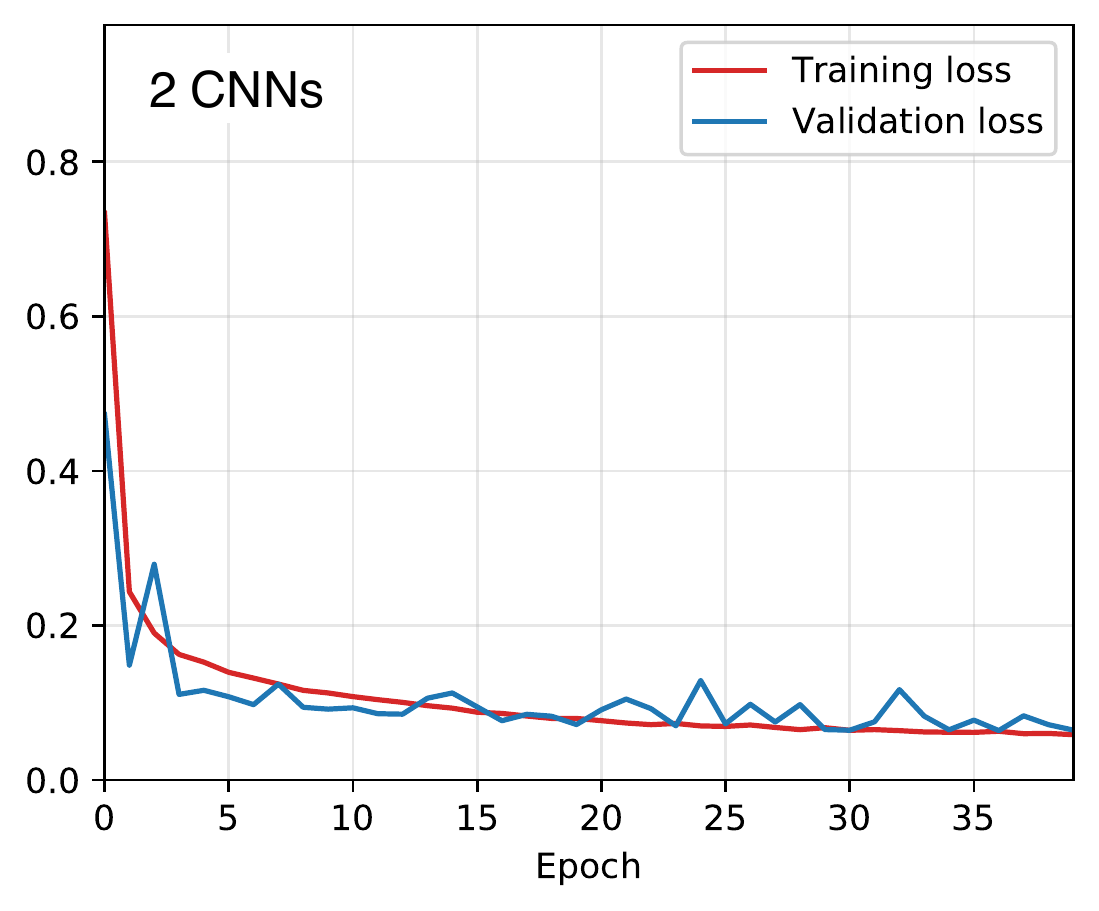}\label{fig:2conv}}
\quad
\sidesubfloat[]{\includegraphics[width=5.0cm]{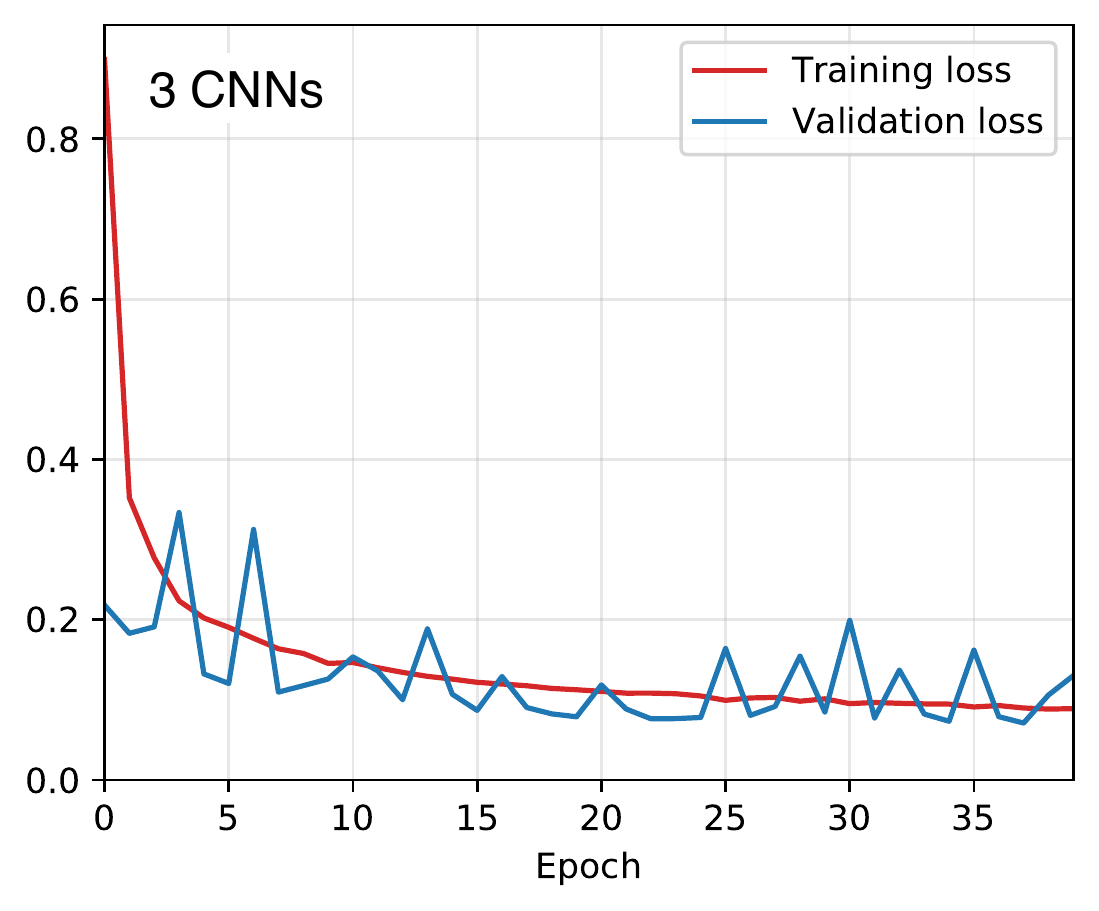}\label{fig:3conv}}
\,
\sidesubfloat[]{\includegraphics[width=5.0cm]{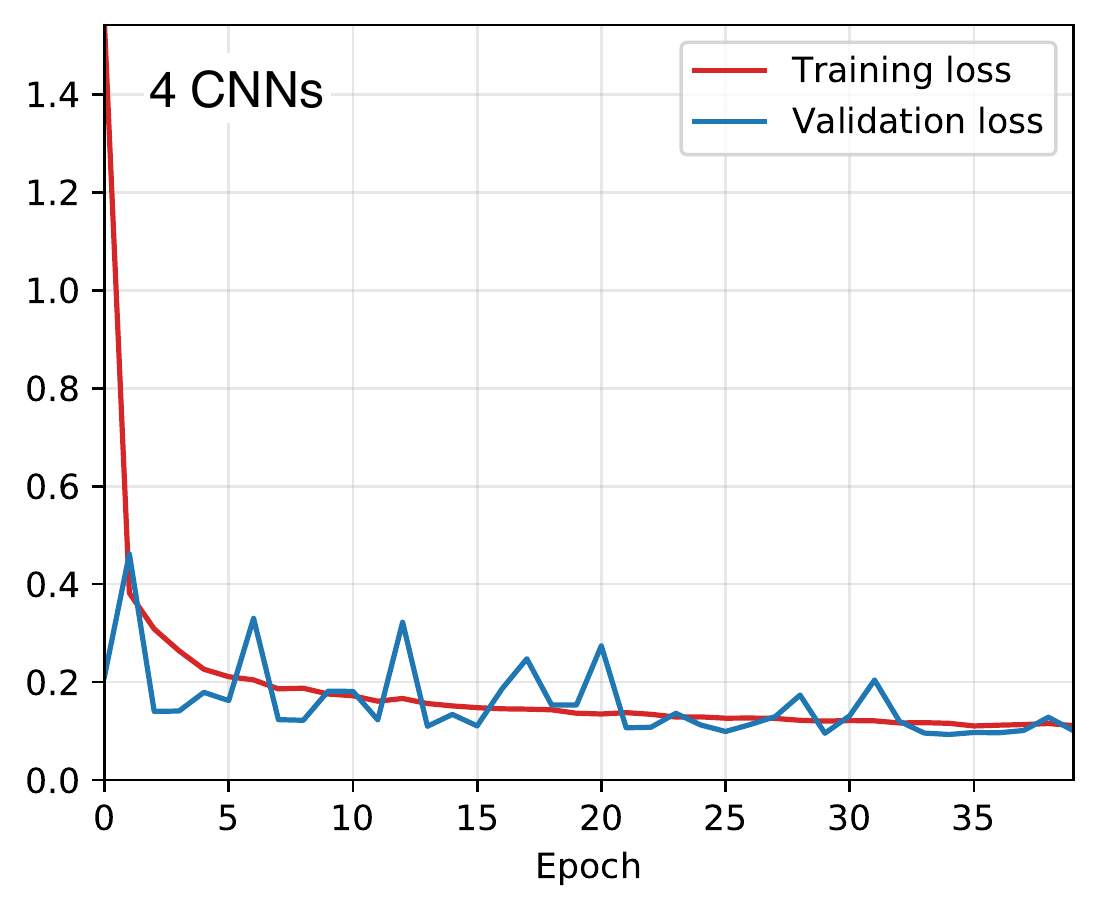}\label{fig:4conv}}
\caption{The training histories of the CNN models with 1,2,3 and 4 convolution layers which corresponds to (a), (b), (c) and (d).}
\label{fig: conv}
\end{figure}

To find a relative optimal CNN model to learn rational parameters from training data-sets, we first examined the performance of different Convolution operations in our CNN models. In the examination, we set 8 consecutive frames as 8 channels per sample, and tested different CNN models containing 1,2,3, and 4 convolution layers. The performances are demonstrated in Fig.~\ref{fig: conv}, in which the training and validation losses (mean square error, MSE) are decreasing with training. In Fig.~\ref{fig:1conv}, the simple CNN model behaves distinct over-fitting after first 5 epochs, which is understandable that the relative concise model tends to over-fit on a large data-set. Although the models with 3 and 4 convolution layers have small training loss as the other models show, their validation losses are highly unstable. It could be interpreted as the lack of training data causes under-fitting. The CNN model with 2 convolution layers are comparatively superior, for its stable performance both on training and validation data-sets. Thus, in following contents, we choose the 2 layer CNN model visualized in Fig.~\ref{fig: cnn} for the further investigations. 

\subsection{Extracting the rational parameters via deep learning}
\begin{figure}[htbp!]
\centering\includegraphics[width=.5\linewidth]{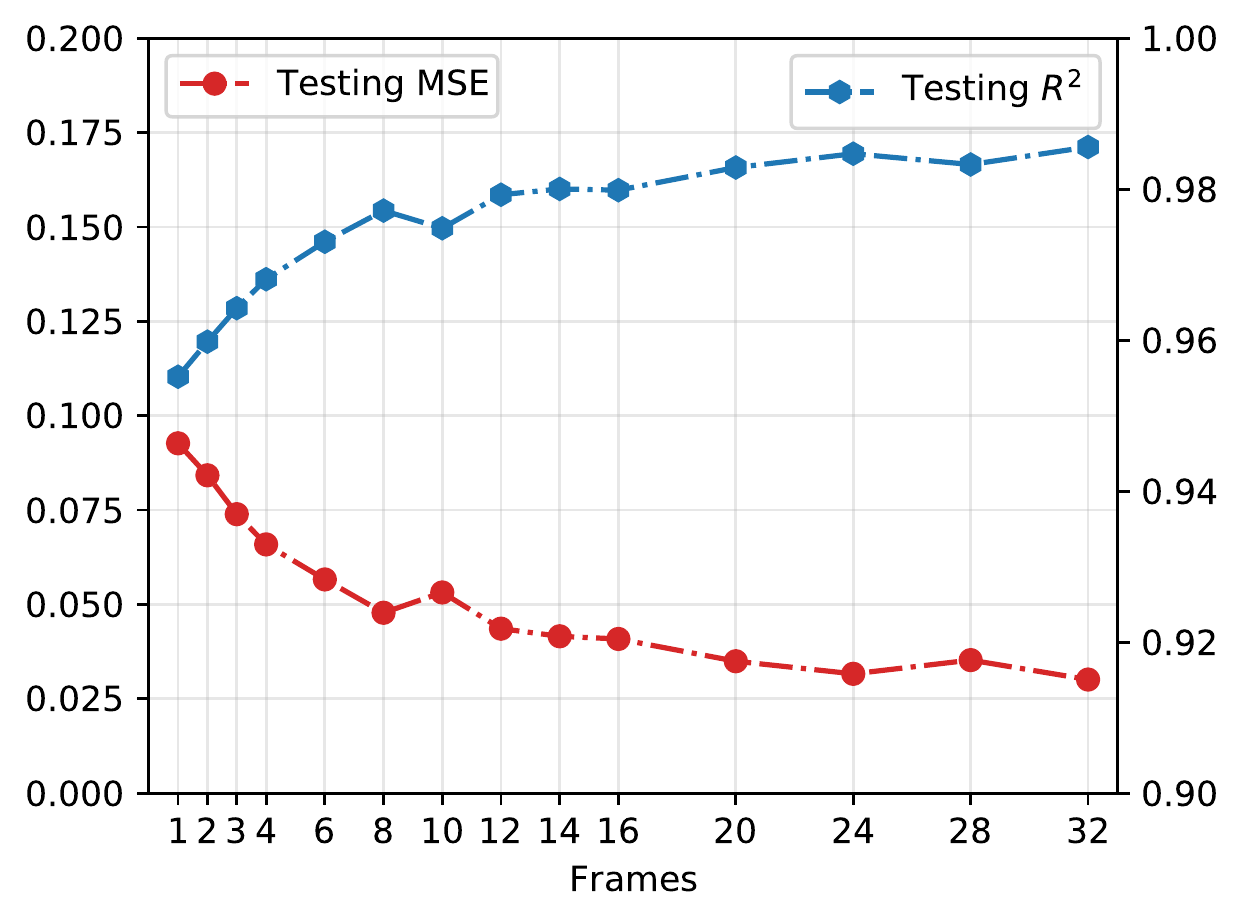}
\caption{Testing performance on different numbers of consecutive frames.}
\label{fig: frames}
\end{figure}
In Fig.~\ref{fig: frames}, we demonstrate the testing performance of the CNN model on different numbers of consecutive frames. The MSE and $R^2 = 1- SS_\mathrm{res}/SS_\mathrm{tot}$ are chosen to evaluate the results learnt from different consecutive evolutions, where $SS_\mathrm{res}$ is the sum of squares of residuals between predictions and ground truths and $SS_\mathrm{tot}$ is the total sum of squares in testing data-set which is proportional to the variance of the data. By increasing the number of frame from 1 to 32, the prediction of rational parameter $\alpha_p$ tends to reach the ground truth $\alpha$. As a relatively ideal result, 8 consecutive frames achieve a testing loss of 0.062 and R-squared value 0.9771. It should be mentioned that while increasing the frame number does increase overall accuracy by a marginal degree, the amount of data (here is time, in real-life applications) required to analyze in these models grows disproportionately against model accuracies. Selecting frame number as low as possible is more realistic for generalizing our framework to assist real-life applications to react more quickly. 

\begin{table}[htbp!]
\centering
\caption{Test accuracy on different initial population densities.}
\label{tab:rho}
\begin{tabular}{@{}lllllll@{}}
\toprule
$\rho_0$ & 0.1          & 0.2           & 0.3           & 0.4           & 0.5           & Mixed \\ \midrule
MSE      & 0.0150   & 0.0397  & 0.0134    & 0.0233   & 0.0095   & 0.0914  \\ \midrule
MAE      & 0.0387   & 0.0609  & 0.0401   & 0.0541   & 0.0289   & 0.2350       \\ \midrule
$R^2$    & 0.9881   & 0.9899  & 0.9910   & 0.9852   & 0.9913   & 0.9565       \\ \bottomrule
\end{tabular}
\end{table}
To analyze the sensitivity of the CNN model to extract rational parameter $\alpha$ under diverse population densities, we tested the CNN model on 5 initial population $\rho_0$ data-sets. To achieve the purpose, we prepare 2000 images from each $\rho_0$ using the same method as previous introduced, but here we feed images from each $\rho_0$ value into the model separately rather than mixed together. 5 well-trained models are tested and shown in 
Table.~\ref{tab:rho}, in which results reveal that for all $\rho_0$ values we examined, the R-squared of the evaluations were all above 0.98, while a 2000-image mixed model gives 0.95. 

\begin{figure}[htbp!]
\centering\includegraphics[width=.5\linewidth]{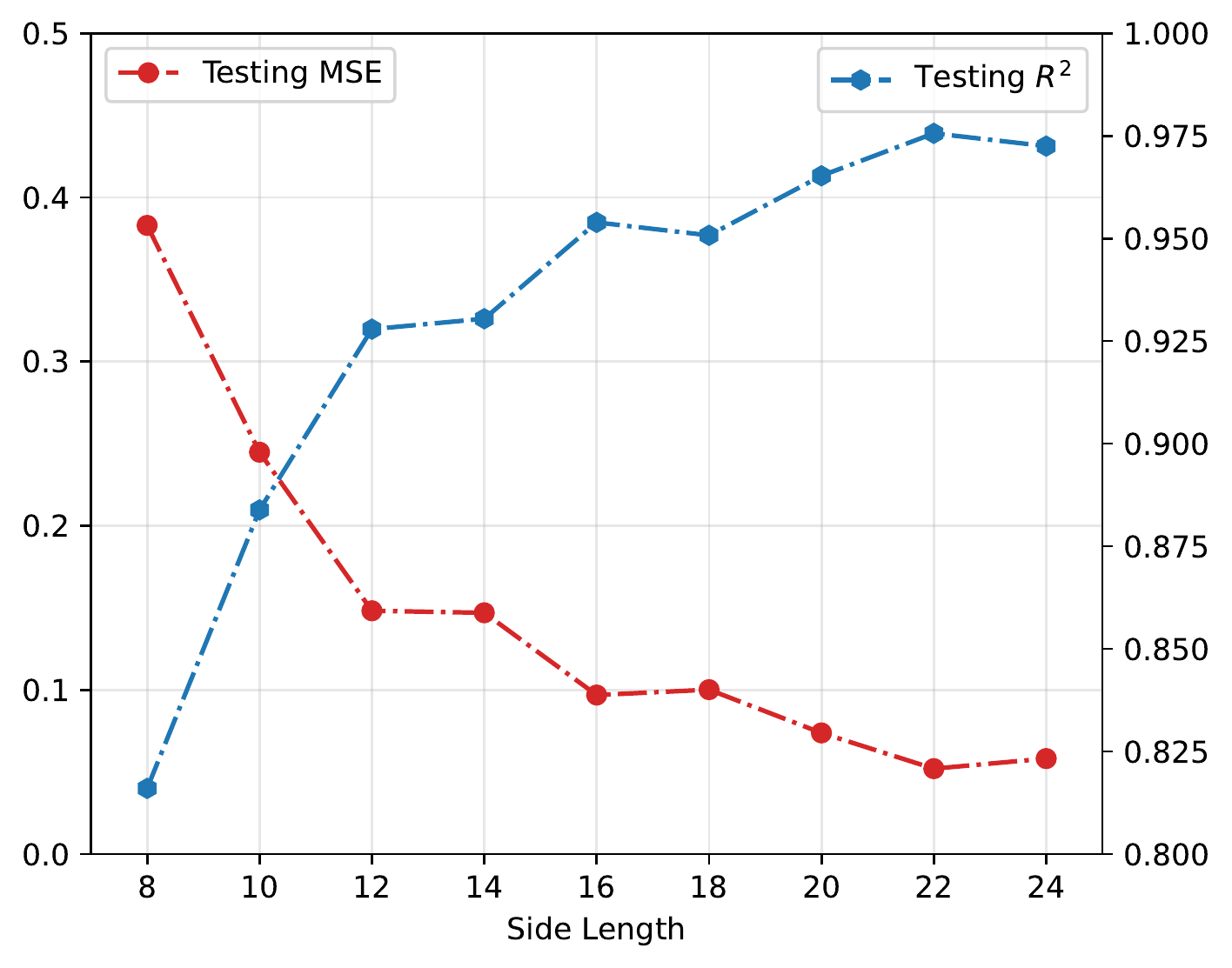}
\caption{Testing performance on different side lengths of images cut from top left corner.}
\label{fig: side}
\end{figure}
Now we are concerned about a more realistic scenario, in which evacuation information is partially missing. The CNN is trained on parts of the images we prepared before. This means that the side length and position of the prepared images are set to be different instead of the numbers of frame. In concise, we select a square area with a given side length off from the 24 by 24 images we generated, and determine the position of the image part by defining the coordinates of the upper left corner on the original image. Using the same 10000 images and 2-layer CNN, the input images are set with side lengths from 8 to 24 \footnote{Images of side lengths less than 8 contain too little information to 
train a 2-layer CNN model}. In Fig.~\ref{fig: side}, results show that the longer the side length, the more accurate the prediction is, which is consist with the information completeness. With regard to the position of cut images, the top left corner and right center (where the exit is) are tested. The inspection using different side lengths as sample shows the obvious advantage of providing information at the outlet. Also, when monitoring the exit, a 12 by 12 image section can achieve an accurate prediction of the entire situation with MSE = 0.094 and $R^2=0.982$, which is close to the performance of training on complete images.

\subsection{Measuring the deviation to the rationality}
\begin{table}[htbp!]
\centering
\caption{Predictions of rational parameters on different initial population densities.}
\label{tab:bounded}
\begin{tabular}{@{}lllllll@{}}
\toprule
$\rho_0$ & 0.1          & 0.2           & 0.3           & 0.4           & 0.5           & Mixed \\ \midrule
$\alpha_\mathrm{p}$    & 2.514   & 2.509  & 2.445   & 2.424   & 2.348   & 2.448 \\\midrule      
$\delta\alpha$    & 0.039   & 0.034  & -0.020   & -0.051   & -0.127   & -0.027 \\  \bottomrule
\end{tabular}
\end{table}
With a well-trained CNN model, we can predict the rational parameter in such evacuation behavior $\alpha$ that reflects the importance of the exit information to individuals. The $ crowd $ rule was introduced to characterize a bounded rational behavior~\cite{wang:2019escape}, in which the deviation of the rationality is measured in our framework. With same processing as Sec.~\ref{sec: ml} to prepare data-sets, we generated 10000 images under the $ crowd $ rule. In a transfer learning manner, the well-trained CNN model learnt on data-set with rational strategy 
is transferred to predict the rational parameters on data-set with the $ crowd $ rule. As Table.~\ref{tab:bounded} shows, predictions of rational parameters on different initial population densities reveal a distinct deviation which is $\delta \alpha$, in which the base-line $\alpha=2.475$. The effect of the $ crowd $ rule to the rational behavior is to reduce the influence of the exit information, or equivalently is to strengthen the importance of the population density in evacuation. With population increasing, the deviation to the rationality  $\delta\alpha$ changes from negative to positive, that is from overestimating rational parameter under small population to underestimating it under large population. In other words, the bounded rationality induced by the $ crowd $ rule in evacuation behavior is quantitatively characterized as the deviation $\delta \alpha$.

\section{Conclusions}
\label{sec: con}
In this study, based on a CA model which generates the spatio-temporal escape maps similar to the actual situation, we propose a deep learning framework to extract the rationality and its deviation induced by heterogeneous information. The latter is introduced in a replicator dynamics describing the bounded decision-making. The well-trained deep CNN accurately predicts the rational factors from multi-frame images generated by the CA model. In addition, it should be noted that the performance of this machine is robust to the incomplete images corresponding to global information loss. This framework provides us with a playground in which the rationality is measured in evacuation and the scheme could also be generalized to other well-designed virtual experiments. It has potentials to be used in recognizing potential collective behavior and avoid trampling on real image data-sets. On the other hand, combining online games with our framework, it would help us to measure the rational degree of group in more general human behaviors~\cite{mao:2016experimental,vandolder:2018wisdom,awad:2018moral,toyokawa:2019social}. In summary, this work provides an insight of quantifying the human rationality with deep learning approaches in collective behaviors.

\section*{Acknowledgments}
The authors thank Yin Jiang and Kai Zhou for useful discussions and comments. The work on this research is supported by the AI grant at FIAS of SAMSON AG, Frankfurt (L. W.).

\bibliographystyle{model1-num-names}
\bibliography{mlevacuation}

\begin{thebibliography}{49}
\expandafter\ifx\csname natexlab\endcsname\relax\def\natexlab#1{#1}\fi
\providecommand{\bibinfo}[2]{#2}
\ifx\xfnm\relax \def\xfnm[#1]{\unskip,\space#1}\fi
\bibitem[{Helbing et~al.(2000)Helbing, Farkas, and
  Vicsek}]{helbing:2000simulating}
\bibinfo{author}{D.~Helbing}, \bibinfo{author}{I.~Farkas},
  \bibinfo{author}{T.~Vicsek},
\newblock \bibinfo{title}{Simulating dynamical features of escape panic},
\newblock \bibinfo{journal}{Nature} \bibinfo{volume}{407}
  (\bibinfo{year}{2000}) \bibinfo{pages}{487--490}.
\bibitem[{Hughes(2002)}]{hughes:2002continuum}
\bibinfo{author}{R.~L. Hughes},
\newblock \bibinfo{title}{A continuum theory for the flow of pedestrians},
\newblock \bibinfo{journal}{Transportation Research Part B: Methodological}
  \bibinfo{volume}{36} (\bibinfo{year}{2002}) \bibinfo{pages}{507--535}.
\bibitem[{Helbing et~al.(2005)Helbing, Buzna, Johansson, and
  Werner}]{helbing:2005selforganized}
\bibinfo{author}{D.~Helbing}, \bibinfo{author}{L.~Buzna},
  \bibinfo{author}{A.~Johansson}, \bibinfo{author}{T.~Werner},
\newblock \bibinfo{title}{Self-{{Organized Pedestrian Crowd Dynamics}}:
  {{Experiments}}, {{Simulations}}, and {{Design Solutions}}},
\newblock \bibinfo{journal}{Transportation Science} \bibinfo{volume}{39}
  (\bibinfo{year}{2005}) \bibinfo{pages}{1--24}.
\bibitem[{Pastor et~al.(2015)Pastor, Garcimart{\'i}n, Gago, Peralta,
  {Mart{\'i}n-G{\'o}mez}, Ferrer, Maza, Parisi, Pugnaloni, and
  Zuriguel}]{pastor:2015experimental}
\bibinfo{author}{J.~M. Pastor}, \bibinfo{author}{A.~Garcimart{\'i}n},
  \bibinfo{author}{P.~A. Gago}, \bibinfo{author}{J.~P. Peralta},
  \bibinfo{author}{C.~{Mart{\'i}n-G{\'o}mez}}, \bibinfo{author}{L.~M. Ferrer},
  \bibinfo{author}{D.~Maza}, \bibinfo{author}{D.~R. Parisi},
  \bibinfo{author}{L.~A. Pugnaloni}, \bibinfo{author}{I.~Zuriguel},
\newblock \bibinfo{title}{Experimental proof of faster-is-slower in systems of
  frictional particles flowing through constrictions},
\newblock \bibinfo{journal}{Phys. Rev. E} \bibinfo{volume}{92}
  (\bibinfo{year}{2015}) \bibinfo{pages}{062817}.
\bibitem[{Nicolas et~al.(2018)Nicolas, Ib{\'a}{\~n}ez, Kuperman, and
  Bouzat}]{nicolas:2018counterintuitive}
\bibinfo{author}{A.~Nicolas}, \bibinfo{author}{S.~Ib{\'a}{\~n}ez},
  \bibinfo{author}{M.~N. Kuperman}, \bibinfo{author}{S.~Bouzat},
\newblock \bibinfo{title}{A counterintuitive way to speed up pedestrian and
  granular bottleneck flows prone to clogging: Can `more' escape faster?},
\newblock \bibinfo{journal}{J. Stat. Mech.} \bibinfo{volume}{2018}
  (\bibinfo{year}{2018}) \bibinfo{pages}{083403}.
\bibitem[{Bain and Bartolo(2019)}]{bain:2019dynamic}
\bibinfo{author}{N.~Bain}, \bibinfo{author}{D.~Bartolo},
\newblock \bibinfo{title}{Dynamic response and hydrodynamics of polarized
  crowds},
\newblock \bibinfo{journal}{Science} \bibinfo{volume}{363}
  (\bibinfo{year}{2019}) \bibinfo{pages}{46--49}.
\bibitem[{Vermuyten et~al.(2016)Vermuyten, Beli{\"e}n, De~Boeck, Reniers, and
  Wauters}]{vermuyten:2016review}
\bibinfo{author}{H.~Vermuyten}, \bibinfo{author}{J.~Beli{\"e}n},
  \bibinfo{author}{L.~De~Boeck}, \bibinfo{author}{G.~Reniers},
  \bibinfo{author}{T.~Wauters},
\newblock \bibinfo{title}{A review of optimisation models for pedestrian
  evacuation and design problems},
\newblock \bibinfo{journal}{Safety Science} \bibinfo{volume}{87}
  (\bibinfo{year}{2016}) \bibinfo{pages}{167--178}.
\bibitem[{Low(2000)}]{low:2000statistical}
\bibinfo{author}{D.~J. Low},
\newblock \bibinfo{title}{Statistical physics: {{Following}} the crowd},
\newblock \bibinfo{journal}{Nature} \bibinfo{volume}{407}
  (\bibinfo{year}{2000}) \bibinfo{pages}{465}.
\bibitem[{Nicolas et~al.(2019)Nicolas, Kuperman, Iba{\~n}ez, Bouzat, and
  {Appert-Rolland}}]{nicolas:2019mechanical}
\bibinfo{author}{A.~Nicolas}, \bibinfo{author}{M.~Kuperman},
  \bibinfo{author}{S.~Iba{\~n}ez}, \bibinfo{author}{S.~Bouzat},
  \bibinfo{author}{C.~{Appert-Rolland}},
\newblock \bibinfo{title}{Mechanical response of dense pedestrian crowds to the
  crossing of intruders},
\newblock \bibinfo{journal}{Sci. Rep.} \bibinfo{volume}{9}
  (\bibinfo{year}{2019}) \bibinfo{pages}{105}.
\bibitem[{Ma et~al.(2021)Ma, Lee, Shi, and Yuen}]{ma:2021spontaneous}
\bibinfo{author}{Y.~Ma}, \bibinfo{author}{E.~W.~M. Lee},
  \bibinfo{author}{M.~Shi}, \bibinfo{author}{R.~K.~K. Yuen},
\newblock \bibinfo{title}{Spontaneous synchronization of motion in pedestrian
  crowds of different densities},
\newblock \bibinfo{journal}{Nat Hum Behav} \bibinfo{volume}{5}
  (\bibinfo{year}{2021}) \bibinfo{pages}{447--457}.
\bibitem[{Helbing and Moln{\'a}r(1995)}]{helbing:1995social}
\bibinfo{author}{D.~Helbing}, \bibinfo{author}{P.~Moln{\'a}r},
\newblock \bibinfo{title}{Social force model for pedestrian dynamics},
\newblock \bibinfo{journal}{Phys. Rev. E} \bibinfo{volume}{51}
  (\bibinfo{year}{1995}) \bibinfo{pages}{4282--4286}.
\bibitem[{Burstedde et~al.(2001)Burstedde, Klauck, Schadschneider, and
  Zittartz}]{burstedde:2001simulation}
\bibinfo{author}{C.~Burstedde}, \bibinfo{author}{K.~Klauck},
  \bibinfo{author}{A.~Schadschneider}, \bibinfo{author}{J.~Zittartz},
\newblock \bibinfo{title}{Simulation of pedestrian dynamics using a
  two-dimensional cellular automaton},
\newblock \bibinfo{journal}{Physica A} \bibinfo{volume}{295}
  (\bibinfo{year}{2001}) \bibinfo{pages}{507--525}.
\bibitem[{Weng et~al.(2006)Weng, Chen, Yuan, and Fan}]{weng:2006cellular}
\bibinfo{author}{W.~G. Weng}, \bibinfo{author}{T.~Chen}, \bibinfo{author}{H.~Y.
  Yuan}, \bibinfo{author}{W.~C. Fan},
\newblock \bibinfo{title}{Cellular automaton simulation of pedestrian counter
  flow with different walk velocities},
\newblock \bibinfo{journal}{Phys. Rev. E} \bibinfo{volume}{74}
  (\bibinfo{year}{2006}) \bibinfo{pages}{036102}.
\bibitem[{Guo et~al.(2012)Guo, Chen, Zheng, and Wei}]{guo:2012heterogeneous}
\bibinfo{author}{X.~Guo}, \bibinfo{author}{J.~Chen},
  \bibinfo{author}{Y.~Zheng}, \bibinfo{author}{J.~Wei},
\newblock \bibinfo{title}{A heterogeneous lattice gas model for simulating
  pedestrian evacuation},
\newblock \bibinfo{journal}{Physica A} \bibinfo{volume}{391}
  (\bibinfo{year}{2012}) \bibinfo{pages}{582--592}.
\bibitem[{Patterson et~al.(2017)Patterson, Fierens, Sangiuliano~Jimka,
  K{\"o}nig, Garcimart{\'i}n, Zuriguel, Pugnaloni, and
  Parisi}]{patterson:2017clogging}
\bibinfo{author}{G.~A. Patterson}, \bibinfo{author}{P.~I. Fierens},
  \bibinfo{author}{F.~Sangiuliano~Jimka}, \bibinfo{author}{P.~G. K{\"o}nig},
  \bibinfo{author}{A.~Garcimart{\'i}n}, \bibinfo{author}{I.~Zuriguel},
  \bibinfo{author}{L.~A. Pugnaloni}, \bibinfo{author}{D.~R. Parisi},
\newblock \bibinfo{title}{Clogging {{Transition}} of {{Vibration}}-{{Driven
  Vehicles Passing}} through {{Constrictions}}},
\newblock \bibinfo{journal}{Phys. Rev. Lett.} \bibinfo{volume}{119}
  (\bibinfo{year}{2017}) \bibinfo{pages}{248301}.
\bibitem[{Aguilar et~al.(2018)Aguilar, Monaenkova, Linevich, Savoie, Dutta,
  Kuan, Betterton, Goodisman, and Goldman}]{aguilar:2018collective}
\bibinfo{author}{J.~Aguilar}, \bibinfo{author}{D.~Monaenkova},
  \bibinfo{author}{V.~Linevich}, \bibinfo{author}{W.~Savoie},
  \bibinfo{author}{B.~Dutta}, \bibinfo{author}{H.-S. Kuan},
  \bibinfo{author}{M.~D. Betterton}, \bibinfo{author}{M.~a.~D. Goodisman},
  \bibinfo{author}{D.~I. Goldman},
\newblock \bibinfo{title}{Collective clog control: {{Optimizing}} traffic flow
  in confined biological and robophysical excavation},
\newblock \bibinfo{journal}{Science} \bibinfo{volume}{361}
  (\bibinfo{year}{2018}) \bibinfo{pages}{672--677}.
\bibitem[{Delarue et~al.(2016)Delarue, Hartung, Schreck, Gniewek, Hu,
  Herminghaus, and Hallatschek}]{delarue:2016selfdriven}
\bibinfo{author}{M.~Delarue}, \bibinfo{author}{J.~Hartung},
  \bibinfo{author}{C.~Schreck}, \bibinfo{author}{P.~Gniewek},
  \bibinfo{author}{L.~Hu}, \bibinfo{author}{S.~Herminghaus},
  \bibinfo{author}{O.~Hallatschek},
\newblock \bibinfo{title}{Self-driven jamming in growing microbial
  populations},
\newblock \bibinfo{journal}{Nat. Phys.} \bibinfo{volume}{12}
  (\bibinfo{year}{2016}) \bibinfo{pages}{762--766}.
\bibitem[{Garcimart{\'i}n et~al.(2015)Garcimart{\'i}n, Pastor, Ferrer, Ramos,
  {Mart{\'i}n-G{\'o}mez}, and Zuriguel}]{garcimartin:2015flow}
\bibinfo{author}{A.~Garcimart{\'i}n}, \bibinfo{author}{J.~M. Pastor},
  \bibinfo{author}{L.~M. Ferrer}, \bibinfo{author}{J.~J. Ramos},
  \bibinfo{author}{C.~{Mart{\'i}n-G{\'o}mez}}, \bibinfo{author}{I.~Zuriguel},
\newblock \bibinfo{title}{Flow and clogging of a sheep herd passing through a
  bottleneck},
\newblock \bibinfo{journal}{Phys. Rev. E} \bibinfo{volume}{91}
  (\bibinfo{year}{2015}) \bibinfo{pages}{022808}.
\bibitem[{Nowak et~al.(2005)Nowak, Vallacher, and
  Zochowski}]{nowak:2005emergence}
\bibinfo{author}{A.~Nowak}, \bibinfo{author}{R.~R. Vallacher},
  \bibinfo{author}{M.~Zochowski},
\newblock \bibinfo{title}{The emergence of personality: {{Dynamic}} foundations
  of individual variation},
\newblock \bibinfo{journal}{Developmental Review} \bibinfo{volume}{25}
  (\bibinfo{year}{2005}) \bibinfo{pages}{351--385}.
\bibitem[{Zanlungo et~al.(2017)Zanlungo, Yucel, Brscic, Kanda, and
  Hagita}]{zanlungo:2017intrinsic}
\bibinfo{author}{F.~Zanlungo}, \bibinfo{author}{Z.~Yucel},
  \bibinfo{author}{D.~Brscic}, \bibinfo{author}{T.~Kanda},
  \bibinfo{author}{N.~Hagita},
\newblock \bibinfo{title}{Intrinsic group behaviour: Dependence of pedestrian
  dyad dynamics on principal social and personal features},
\newblock \bibinfo{journal}{PLOS ONE} \bibinfo{volume}{12}
  (\bibinfo{year}{2017}) \bibinfo{pages}{e0187253}.
\bibitem[{Corbetta et~al.(2017)Corbetta, Lee, Benzi, Muntean, and
  Toschi}]{corbetta:2017fluctuations}
\bibinfo{author}{A.~Corbetta}, \bibinfo{author}{C.-m. Lee},
  \bibinfo{author}{R.~Benzi}, \bibinfo{author}{A.~Muntean},
  \bibinfo{author}{F.~Toschi},
\newblock \bibinfo{title}{Fluctuations around mean walking behaviors in diluted
  pedestrian flows},
\newblock \bibinfo{journal}{Phys. Rev. E} \bibinfo{volume}{95}
  (\bibinfo{year}{2017}) \bibinfo{pages}{032316}.
\bibitem[{Nicolas et~al.(2018)Nicolas, Garcimart{\'i}n, and
  Zuriguel}]{nicolas:2018trap}
\bibinfo{author}{A.~Nicolas}, \bibinfo{author}{{\'A}.~Garcimart{\'i}n},
  \bibinfo{author}{I.~Zuriguel},
\newblock \bibinfo{title}{Trap {{Model}} for {{Clogging}} and {{Unclogging}} in
  {{Granular Hopper Flows}}},
\newblock \bibinfo{journal}{Phys. Rev. Lett.} \bibinfo{volume}{120}
  (\bibinfo{year}{2018}) \bibinfo{pages}{198002}.
\bibitem[{Cavagna et~al.(2018)Cavagna, Giardina, and
  Grigera}]{cavagna:2018physics}
\bibinfo{author}{A.~Cavagna}, \bibinfo{author}{I.~Giardina},
  \bibinfo{author}{T.~S. Grigera},
\newblock \bibinfo{title}{The physics of flocking: {{Correlation}} as a compass
  from experiments to theory},
\newblock \bibinfo{journal}{Physics Reports} \bibinfo{volume}{728}
  (\bibinfo{year}{2018}) \bibinfo{pages}{1--62}.
\bibitem[{Castellano et~al.(2009)Castellano, Fortunato, and
  Loreto}]{castellano:2009statistical}
\bibinfo{author}{C.~Castellano}, \bibinfo{author}{S.~Fortunato},
  \bibinfo{author}{V.~Loreto},
\newblock \bibinfo{title}{Statistical physics of social dynamics},
\newblock \bibinfo{journal}{Rev. Mod. Phys.} \bibinfo{volume}{81}
  (\bibinfo{year}{2009}) \bibinfo{pages}{591--646}.
\bibitem[{Ball(2012)}]{ball:2012why}
\bibinfo{author}{P.~Ball}, \bibinfo{title}{Why {{Society}} Is a {{Complex
  Matter}}: {{Meeting Twenty}}-First {{Century Challenges}} with a {{New Kind}}
  of {{Science}}}, \bibinfo{publisher}{{Springer}}, \bibinfo{address}{{Berlin
  Heidelberg}}, \bibinfo{year}{2012}.
\bibitem[{Moussa{\"i}d et~al.(2011)Moussa{\"i}d, Helbing, and
  Theraulaz}]{moussaid:2011how}
\bibinfo{author}{M.~Moussa{\"i}d}, \bibinfo{author}{D.~Helbing},
  \bibinfo{author}{G.~Theraulaz},
\newblock \bibinfo{title}{How simple rules determine pedestrian behavior and
  crowd disasters},
\newblock \bibinfo{journal}{PNAS} \bibinfo{volume}{108} (\bibinfo{year}{2011})
  \bibinfo{pages}{6884--6888}.
\bibitem[{Helbing et~al.(2011)Helbing, Tr{\"o}ster, Wirz, and
  Roggen}]{helbing:2011recognition}
\bibinfo{author}{D.~Helbing}, \bibinfo{author}{G.~Tr{\"o}ster},
  \bibinfo{author}{M.~Wirz}, \bibinfo{author}{D.~Roggen},
\newblock \bibinfo{title}{Recognition of crowd behavior from mobile sensors
  with pattern analysis and graph clustering methods},
\newblock \bibinfo{journal}{Netw. Heterog. Media} \bibinfo{volume}{6}
  (\bibinfo{year}{2011}) \bibinfo{pages}{521--544}.
\bibitem[{Wang and Weng(2018)}]{wang:2018study}
\bibinfo{author}{C.~Wang}, \bibinfo{author}{W.~Weng},
\newblock \bibinfo{title}{Study on the collision dynamics and the transmission
  pattern between pedestrians along the queue},
\newblock \bibinfo{journal}{J. Stat. Mech.} \bibinfo{volume}{2018}
  (\bibinfo{year}{2018}) \bibinfo{pages}{073406}.
\bibitem[{Rahman and Hasan(2018)}]{rahman:2018shortterm}
\bibinfo{author}{R.~Rahman}, \bibinfo{author}{S.~Hasan},
\newblock \bibinfo{title}{Short-{{Term Traffic Speed Prediction}} for
  {{Freeways During Hurricane Evacuation}}: {{A Deep Learning Approach}}},
\newblock in: \bibinfo{booktitle}{2018 21st {{International Conference}} on
  {{Intelligent Transportation Systems}} ({{ITSC}})}, pp.
  \bibinfo{pages}{1291--1296}.
\bibitem[{Song et~al.(2017)Song, Shibasaki, Yuan, Xie, Li, and
  Adachi}]{song:2017deepmob}
\bibinfo{author}{X.~Song}, \bibinfo{author}{R.~Shibasaki},
  \bibinfo{author}{N.~J. Yuan}, \bibinfo{author}{X.~Xie},
  \bibinfo{author}{T.~Li}, \bibinfo{author}{R.~Adachi},
\newblock \bibinfo{title}{{{DeepMob}}: {{Learning Deep Knowledge}} of {{Human
  Emergency Behavior}} and {{Mobility}} from {{Big}} and {{Heterogeneous
  Data}}},
\newblock \bibinfo{journal}{ACM Trans. Inf. Syst.} \bibinfo{volume}{35}
  (\bibinfo{year}{2017}) \bibinfo{pages}{41:1--41:19}.
\bibitem[{Chen et~al.(2020)Chen, Hu, Mao, Deng, and Gao}]{chen:2020application}
\bibinfo{author}{Y.~Chen}, \bibinfo{author}{S.~Hu}, \bibinfo{author}{H.~Mao},
  \bibinfo{author}{W.~Deng}, \bibinfo{author}{X.~Gao},
\newblock \bibinfo{title}{Application of the best evacuation model of deep
  learning in the design of public structures},
\newblock \bibinfo{journal}{Image and Vision Computing} \bibinfo{volume}{102}
  (\bibinfo{year}{2020}) \bibinfo{pages}{103975}.
\bibitem[{Pang et~al.(2018)Pang, Zhou, Su, Petersen, St{\"o}cker, and
  Wang}]{pang:2018equationofstatemeter}
\bibinfo{author}{L.-G. Pang}, \bibinfo{author}{K.~Zhou},
  \bibinfo{author}{N.~Su}, \bibinfo{author}{H.~Petersen},
  \bibinfo{author}{H.~St{\"o}cker}, \bibinfo{author}{X.-N. Wang},
\newblock \bibinfo{title}{An equation-of-state-meter of quantum chromodynamics
  transition from deep learning},
\newblock \bibinfo{journal}{Nature Commun.} \bibinfo{volume}{9}
  (\bibinfo{year}{2018}) \bibinfo{pages}{210}.
\bibitem[{Jiang et~al.(2021)Jiang, Wang, and Zhou}]{jiang:2021deep}
\bibinfo{author}{L.~Jiang}, \bibinfo{author}{L.~Wang},
  \bibinfo{author}{K.~Zhou},
\newblock \bibinfo{title}{Deep learning stochastic processes with {{QCD}} phase
  transition},
\newblock \bibinfo{journal}{ArXiv210304090 Nucl-Th Physicsphysics}
  (\bibinfo{year}{2021}).
\bibitem[{Wang et~al.(2021)Wang, Xu, Stoecker, Stoecker, Jiang, and
  Zhou}]{wang:2021machine}
\bibinfo{author}{L.~Wang}, \bibinfo{author}{T.~Xu},
  \bibinfo{author}{T.~Stoecker}, \bibinfo{author}{H.~Stoecker},
  \bibinfo{author}{Y.~Jiang}, \bibinfo{author}{K.~Zhou},
\newblock \bibinfo{title}{Machine learning spatio-temporal epidemiological
  model to evaluate {{Germany}}-county-level {{COVID}}-19 risk},
\newblock \bibinfo{journal}{Mach. Learn.: Sci. Technol.}
  (\bibinfo{year}{2021}).
\bibitem[{Wang and Jiang(2019)}]{wang:2019escape}
\bibinfo{author}{L.~Wang}, \bibinfo{author}{Y.~Jiang},
\newblock \bibinfo{title}{Escape dynamics based on bounded rationality},
\newblock \bibinfo{journal}{Physica A} \bibinfo{volume}{531}
  (\bibinfo{year}{2019}) \bibinfo{pages}{121777}.
\bibitem[{Pan et~al.(2014)Pan, Wang, Shi, Wang, and He}]{pan:2014spatial}
\bibinfo{author}{Q.~Pan}, \bibinfo{author}{L.~Wang}, \bibinfo{author}{R.~Shi},
  \bibinfo{author}{H.~Wang}, \bibinfo{author}{M.~He},
\newblock \bibinfo{title}{Spatial modes of cooperation based on bounded
  rationality},
\newblock \bibinfo{journal}{Physica A} \bibinfo{volume}{415}
  (\bibinfo{year}{2014}) \bibinfo{pages}{421--427}.
\bibitem[{Heli{\"o}vaara et~al.(2013)Heli{\"o}vaara, Ehtamo, Helbing, and
  Korhonen}]{heliovaara:2013patient}
\bibinfo{author}{S.~Heli{\"o}vaara}, \bibinfo{author}{H.~Ehtamo},
  \bibinfo{author}{D.~Helbing}, \bibinfo{author}{T.~Korhonen},
\newblock \bibinfo{title}{Patient and impatient pedestrians in a spatial game
  for egress congestion},
\newblock \bibinfo{journal}{Phys. Rev. E} \bibinfo{volume}{87}
  (\bibinfo{year}{2013}) \bibinfo{pages}{012802}.
\bibitem[{Taylor and Jonker(1978)}]{taylor:1978evolutionary}
\bibinfo{author}{P.~D. Taylor}, \bibinfo{author}{L.~B. Jonker},
\newblock \bibinfo{title}{Evolutionary stable strategies and game dynamics},
\newblock \bibinfo{journal}{Mathematical Biosciences} \bibinfo{volume}{40}
  (\bibinfo{year}{1978}) \bibinfo{pages}{145--156}.
\bibitem[{Simon(1983)}]{simon:1983reason}
\bibinfo{author}{H.~A. Simon}, \bibinfo{title}{Reason in Human Affairs},
  \bibinfo{publisher}{{Stanford Univ. Press}}, \bibinfo{address}{{Stanford,
  Calif}}, \bibinfo{year}{1983}.
\bibitem[{Gigerenzer and Selten(2002)}]{gigerenzer:2002bounded}
\bibinfo{author}{G.~Gigerenzer}, \bibinfo{author}{R.~Selten},
  \bibinfo{title}{Bounded {{Rationality}}: {{The Adaptive Toolbox}}},
  \bibinfo{publisher}{{MIT Press}}, \bibinfo{year}{2002}.
\bibitem[{Yang and Yao(2005)}]{yang:2005walrasian}
\bibinfo{author}{X.~Yang}, \bibinfo{author}{S.~Yao},
\newblock \bibinfo{title}{Walrasian sequential equilibrium, bounded
  rationality, and social experiments},
\newblock \bibinfo{journal}{Div. Labor Transaction Costs} \bibinfo{volume}{01}
  (\bibinfo{year}{2005}) \bibinfo{pages}{73--98}.
\bibitem[{B{\u a}beanu and Garlaschelli(2018)}]{babeanu:2018evidence}
\bibinfo{author}{A.-I. B{\u a}beanu}, \bibinfo{author}{D.~Garlaschelli},
\newblock \bibinfo{title}{Evidence for {{Mixed Rationalities}} in {{Preference
  Formation}}},
\newblock \bibinfo{journal}{Complexity} \bibinfo{volume}{2018}
  (\bibinfo{year}{2018}) \bibinfo{pages}{1--19}.
\bibitem[{Lee et~al.(2001)Lee, Hui, Wang, and Johnson}]{lee:2001effects}
\bibinfo{author}{K.~Lee}, \bibinfo{author}{P.~M. Hui}, \bibinfo{author}{B.-H.
  Wang}, \bibinfo{author}{N.~F. Johnson},
\newblock \bibinfo{title}{Effects of {{Announcing Global Information}} in a
  {{Two}}-{{Route Traffic Flow Model}}},
\newblock \bibinfo{journal}{J. Phys. Soc. Jpn.} \bibinfo{volume}{70}
  (\bibinfo{year}{2001}) \bibinfo{pages}{3507--3510}.
\bibitem[{Wang et~al.(2005)Wang, Wang, Zheng, Yin, and
  Zhou}]{wang:2005advanced}
\bibinfo{author}{W.-X. Wang}, \bibinfo{author}{B.-H. Wang},
  \bibinfo{author}{W.-C. Zheng}, \bibinfo{author}{C.-Y. Yin},
  \bibinfo{author}{T.~Zhou},
\newblock \bibinfo{title}{Advanced information feedback in intelligent traffic
  systems},
\newblock \bibinfo{journal}{Phys. Rev. E} \bibinfo{volume}{72}
  (\bibinfo{year}{2005}) \bibinfo{pages}{066702}.
\bibitem[{Kirchner et~al.(2003)Kirchner, Nishinari, and
  Schadschneider}]{kirchner:2003friction}
\bibinfo{author}{A.~Kirchner}, \bibinfo{author}{K.~Nishinari},
  \bibinfo{author}{A.~Schadschneider},
\newblock \bibinfo{title}{Friction effects and clogging in a cellular automaton
  model for pedestrian dynamics},
\newblock \bibinfo{journal}{Phys. Rev. E} \bibinfo{volume}{67}
  (\bibinfo{year}{2003}) \bibinfo{pages}{056122}.
\bibitem[{Mao et~al.(2016)Mao, Mason, Suri, and Watts}]{mao:2016experimental}
\bibinfo{author}{A.~Mao}, \bibinfo{author}{W.~Mason},
  \bibinfo{author}{S.~Suri}, \bibinfo{author}{D.~J. Watts},
\newblock \bibinfo{title}{An {{Experimental Study}} of {{Team Size}} and
  {{Performance}} on a {{Complex Task}}},
\newblock \bibinfo{journal}{PLOS ONE} \bibinfo{volume}{11}
  (\bibinfo{year}{2016}) \bibinfo{pages}{e0153048}.
\bibitem[{{van Dolder} and {van den Assem}(2018)}]{vandolder:2018wisdom}
\bibinfo{author}{D.~{van Dolder}}, \bibinfo{author}{M.~J. {van den Assem}},
\newblock \bibinfo{title}{The wisdom of the inner crowd in three large natural
  experiments},
\newblock \bibinfo{journal}{Nat Hum Behav} \bibinfo{volume}{2}
  (\bibinfo{year}{2018}) \bibinfo{pages}{21--26}.
\bibitem[{Awad et~al.(2018)Awad, Dsouza, Kim, Schulz, Henrich, Shariff,
  Bonnefon, and Rahwan}]{awad:2018moral}
\bibinfo{author}{E.~Awad}, \bibinfo{author}{S.~Dsouza},
  \bibinfo{author}{R.~Kim}, \bibinfo{author}{J.~Schulz},
  \bibinfo{author}{J.~Henrich}, \bibinfo{author}{A.~Shariff},
  \bibinfo{author}{J.-F. Bonnefon}, \bibinfo{author}{I.~Rahwan},
\newblock \bibinfo{title}{The {{Moral Machine}} experiment},
\newblock \bibinfo{journal}{Nature} \bibinfo{volume}{563}
  (\bibinfo{year}{2018}) \bibinfo{pages}{59}.
\bibitem[{Toyokawa et~al.(2019)Toyokawa, Whalen, and
  Laland}]{toyokawa:2019social}
\bibinfo{author}{W.~Toyokawa}, \bibinfo{author}{A.~Whalen},
  \bibinfo{author}{K.~N. Laland},
\newblock \bibinfo{title}{Social learning strategies regulate the wisdom and
  madness of interactive crowds},
\newblock \bibinfo{journal}{Nat. Hum. Behav.} \bibinfo{volume}{3}
  (\bibinfo{year}{2019}) \bibinfo{pages}{183--193}.

\end{thebibliography}







\end{document}